\documentstyle[rotating,12pt]{article}
\def\baselinestretch{1.8}
\newcommand\kappatwid{\widetilde \kappa}

\newcommand\lambdatwid{\widetilde \lambda}
\def \p {\mbox{\boldmath $p$}}

\begin{document}
%\begin{titlepage}
%\vspace*{0.5truein}

\begin{flushright}
\begin{tabular}{l}
%hep-ph/9411324 \\
BU-TH-94/06  \\
MRI-PHY/16/94 \\
LNF-94/067 P \\
%July, 1994
\end{tabular}
\end{flushright}
\vskip .6cm

\begin{center}
{\large\bf TRIPLET HIGGS BOSONS AT $e^+ e^-$ COLLIDERS\\[0.5truein]}
{ Rohini Godbole$^1$\\}
Physics Department, University of Bombay, Vidyanagari, Bombay - 400 098,
India \\
{ Biswarup Mukhopadhyaya$^2$\\}
Mehta Research Institute, 10 Kasturba Gandhi Marg, Allahabad - 211 002, India
\\
{ Marek Nowakowski$^3$\\}
INFN- Laboratori Nazionali di Frascati, P.O.Box 13, I-00044, Frascati(Roma),
Italy \\
\end{center}
\vskip .5cm

\centerline{\bf ABSTRACT}
We investigate the possibility of probing the  triplet Higgs boson sector
via single charged Higgs production in the process  $e^+ e^- \to H^+ l \nu_l$,
 at high energy electron-positron colliders,
using  the tree level $H^+W^-Z$ coupling which is a
unique feature of such models. We find that even LEP-200 can give nontrivial
information upto $M_{H^{\pm}} \simeq 120 $ GeV if the doublet -triplet
mixing is not restricted by the current value of the $\rho$ parameter which is
the case in models with a custodial symmetry. Further we point out that in such
models,  the 4-body, tree level decay $H^+ \to W^*Z^* \to 4$
fermions  dominates and hence provides a very clean signal when the four
fermions are leptons. At NLC  the discovery range for the
charged  Higgs in triplet models via this process is $\sim 400$ GeV.

%\hspace*{\fill}

\noindent

E-mail : $^{1}rohini@theory.tifr.res.in  \;\; ^{2}
biswarup@mri.ernet.in \;\; ^{3}nowakowski@lnf.infn.it$

%\end{titlepage}
\newpage
\textheight=8.9in

   There is no evidence as yet of the existence of the scalar particle(s)
which are essential for the spontaneous breakdown of the
$SU(2)_L \otimes U(1)_Y$ electroweak symmetry. Although a single scalar $SU(2)$
doublet suffices for the purpose, it is by no means compulsive for
elementary scalars, if they exist at all in nature, to be restricted to
one doublet only. Theories with two or more doublets are often explored in
this spirit \cite{2hd}. Also, it is a tenable hypothesis that there are not
only doublets
but also higher representations of $SU(2)$, such as triplets, comprising
scalar particles \cite{tri}. It is particularly interesting that whereas all
the
charged fermions must get their masses via Yukawa couplings with Higgs
doublets, the vacuum expectation  values (vev) of triplets can give
masses to neutrinos without requiring any  "sterile" right-handed neutrino
species. Since the existence of nonzero neutrino masses is suggested by
phenomena such as the solar neutrino puzzle \cite{solar}, it might be useful to
consider this possibility of a different origin of neutrino masses,
which in turn might help us understand why they are so small compared
to those of the other fermions.

  However, one is faced with the problem that the vev of the neutral
member of the triplet gives additional contributions to the parameter
$\rho = m_W^2/(m_Z^2 \cos^2 \theta_W)$ (where $\theta_W$ is the Weinberg
angle) at the tree-level, tending to change its value of unity which is
guaranteed if the  representation of the Higgs is restricted to only doublets.
Since the present experimental value of $\rho$ is $(1.0004 \pm 0.0022
 \pm 0.002)$ \cite{rho}, any scenario with scalar  triplets must be
constrained accordingly.

 One option to build this constraint into the model is to postulate that the
vev of the neutral member of the triplet is small enough so that its
contribution to $\rho$ is within the experimental limits. This smallness
can then be translated into an upper limit on the mixing angle between the
doublet and the triplet.

   The other option is to make the ingenious asumption, first suggested by
Georgi and Machacek \cite{georgi} and by Chanowitz and Golden \cite{chano},
that there are in fact more than one triplets, arranged
in such a manner that their contributions to the $\rho$-parameer cancel
each other. In order that this may happen, one needs to have one complex
triplet $\Delta$ (with hypercharge Y=2)
and one real $\chi$ (Y=0) triplet, in addition to the Y=1 complex doublet of
the
minimal standard model. Furthermore the vev's of these two triplet fields
must be equal to guarantee $\rho=1$. Therefore, there is no restriction on
the vev of the triplet  and hence on the doublet-triplet mixing.

    In this paper, we examine some testable consequences of the existence of
triplets in both these situations. Several such studies have already
been done in recent times in the context of $e^+ e^-$ as well as of hadronic
colliders \cite{dicus}, \cite{gunion}. However, we focus here on a specific
interaction which is a distinguishing feature of triplet scalars
(or those of higher representations), namely, a tree-level interaction
involving the W, the Z and a charged scalar and which has not been studied
in detail before. This interaction vertex is absent
with an arbitrary number of Higgs doublets added to the particle
spectrum (for loop induced $H^+W^-Z$ vertex see \cite{hwz}). The
possibility of constraining the triplet sector using the above
tree level coupling in Z-factories has been discussed earlier by one of
the present authors \cite{bisw}. Here we investigate its
observable consequences in LEP-II and more energetic $e^+e^-$ colliders. As
we shall see below, it leads to some very characteristic signals which are
considerably free from standard model backgrounds.

 Let us first summarize the main features of the scalar sector for the two
possibilities mentioned above.
 With a complex triplet ($Y=2$) $\Delta$ and a doublet $\phi$ given by
\begin{equation} \label{e1}
\Delta=\left( \begin{array}{c}
\Delta^{++} \\
\Delta^+ \\
\Delta^0
\end{array} \right),\, \, \, \, \,
\phi=\left( \begin{array}{c}
\phi^+ \\
\phi^0
\end{array}\right)
\end{equation}
the most general Higgs potential including $\Delta$ and $\phi$  reads
\begin{eqnarray} \label{e2}
V_{\phi \Delta}&=& \mu_1^2(\phi^{\dagger}\phi)+\mu_2^2 (\Delta^{\dagger} \Delta
)+ \lambda_1 (\phi^{\dagger} \phi)^2 + \lambda_2 (\Delta^{\dagger}\Delta)^2
+ \lambda_3 |\Delta^T C' \Delta |^2 + \lambda_4(\phi^{\dagger} \phi)(\Delta
^{\dagger} \Delta) \nonumber \\
&+& \lambda_5 (\Delta^{\dagger} T_i \Delta )(\phi^{\dagger} {\tau_i \over 2}
\phi) -i\lambda_6 \left(\Delta_P^{i*}(\phi^T \tau_2 {\tau_i \over 2}\phi) +{\rm
h.c.}
\right)
\end{eqnarray}
where $\Delta_P=P^{\dagger}\Delta$ and $P$ is a unitary matrix which relates
the real 3-dimensional $SU(2)$ representation with generators
$(\tilde{T}_j)_{ik}
=i\epsilon_{ijk}$ to the equivalent representation given by $T_i=P\tilde{T}_i
P^{\dagger}$ with diagonal $T_3$. The matrices $P$ and $C'$ are then
\begin{equation} \label{e3}
P=\left( \begin{array}{ccc}
-1/\sqrt{2} & i/\sqrt{2} & 0  \\
0 & 0 & 1  \\
1/\sqrt{2} & i/\sqrt{2} & 0
\end{array}\right),
\, \, \, \, \, \,
C'=P^* P^{\dagger}=\left( \begin{array}{ccc}
0 & 0 & -1  \\
0 & 1 & 0  \\
-1 & 0 & 0
\end{array}\right)
\end{equation}
Note here that the potential of eq. (\ref{e2}) breaks lepton number explictly
and the model therefore contains no
majoron in the physical particle spectrum \cite{ronc} (a triplet majoron is
now excluded by LEP measurements). Taking the vacuum expectation values
to be $<\phi^0>=v$ and $<\Delta^0>=w$ the $\rho$ parameter in this model is
$\rho=(1+{2w^2 \over v^2})/(1+ {4w^2 \over v^2})$. The current experimental
constraint mentioned above translates (at $99\%$ confidence level) to
\begin{equation} \label{e4}
%{w \over v}\stackrel{<}{\sim} 0.066
{w \over v} \leq 0.066
\end{equation}
The mass eigenstates for the charged scalars can be obtained from
\begin{equation} \label{e5}
\left(\begin{array}{c}
H^{'+}\\
G^+
\end{array}\right)
= \left(\begin{array}{cc}
-s_{H'} & c_{H'} \\
c_{H'} & s_{H'}
\end{array}\right) \left(\begin{array}{c}
\phi^+ \\
\Delta^+
\end{array}\right)
\end{equation}
where $G^+$ is the would-be Goldstone boson. The mixing angles are ($s_{H'}
\equiv \sin \theta_{H'}$ etc.)
\begin{equation} \label{e6}
s_{H'}={\sqrt{2} w \over \sqrt{v^2+2w^2}}, \, \, \,\, c_{H'}={v \over
\sqrt{v^2 + 2w^2}}
\end{equation}
It is worth noting here that  in this case the $H^{'+}$ has
tree level couplings to fermions through the  doublet component. The mass
of the charged boson is given by
\begin{equation} \label{e7}
M_{H^{'+}}^2=-({\lambda_5 \over 2}+ {\lambda_6 \over \sqrt{2}w})(v^2+2w^2)
\end{equation}
Since two scales , $v$ and $w$, are involved the dimensional
coupling constant $\lambda_6$ can be either ${\cal O}(1)w$ or ${\cal O}(1)v$.
In the former case we expect the charged Higgs boson mass to be $\sim {\cal O}
(v^2)$ whereas in the latter case this charged particle can be much heavier
$\sim {\cal O}({v \over w}v^2)$.

It is known that one has to worry about hierarchy problems because of the large
splitting of the two vacuum expectation values implied by eq. (\ref{e4}).
One way to avoid this problem would be to supersymmetrize the model (for
supersymmetric model with a real ($Y=0$) triplet see \cite{stri}).
While discussing below the phenomenological consequences of triplet models we
will consider the model given by eq. (\ref{e2}) as a representative of
models with small triplet vev (e.g. a supersymmetric version of the same).
Yet another solution to the above mentioned problem suggested
by the authors of refs. \cite{georgi}, \cite{chano} is to add a real triplet
field $\chi$ ($Y=0$)
\begin{equation} \label{e8}
\chi=\left(\begin{array}{c}
\chi^+ \\
\chi^0 \\
\chi^-
\end{array}\right)
\end{equation}
and impose on the Higgs potential a $SU(2)_L \otimes SU(2)_R$ symmetry which
at tree level
forces the two vacuum expectation values, $<\Delta^0>$ and $<\chi^0>$ to be
equal. As a result the $\rho$ parameter is one at tree level. The fields
carrying two group indices can be conveniently represented by
\begin{equation} \label{e9}
\Phi=\left(\begin{array}{cc}
\phi^{0*} & \phi^+ \\
\phi^- & \phi^0
\end{array}\right),\, \, \, \, \,
\xi=\left(\begin{array}{ccc}
\Delta^0 & \chi^+ & \Delta^{++} \\
\Delta^- & \chi^0 & \Delta^+ \\
\Delta^{--} & \chi^- & \Delta^{0*}
\end{array}\right)
\end{equation}
The corresponding Higgs potential is then
\begin{eqnarray} \label{e10}
V_{\phi \Delta \chi}&=&\tilde{\mu}_1^2{\rm Tr}(\Phi^{\dagger}\Phi) +
\tilde{\mu}_2^2{\rm Tr}(\xi^{\dagger}\xi)+ \tilde{\lambda}_1{\rm Tr}(\Phi^
{\dagger}\Phi)^2 + \tilde{\lambda}_2{\rm Tr}(\xi^{\dagger}\xi)^2 +
\tilde{\lambda}_3{\rm Tr}(\Phi^{\dagger}\Phi){\rm Tr}(\xi^{\dagger}\xi)
\nonumber \\
&+& \tilde{\lambda}_4{\rm Tr}(\xi^{\dagger}\xi \xi^{\dagger}\xi) +
\tilde{\lambda}_5{\rm Tr}(\Phi^{\dagger}{\tau_i \over 2}\Phi {\tau_j \over 2})
{\rm Tr}(\xi^{\dagger}T_i \xi T_j)\nonumber \\
&+&  \tilde{\lambda}_6{\rm Tr}(\Phi^{\dagger}{\tau_i \over 2}\Phi {\tau_j
\over 2})\xi^{ij}_P + \tilde{\lambda}_7{\rm Tr}(\xi^{\dagger}T_i \xi T_j)
\xi^{ij}_P
\end{eqnarray}
where $\xi_P=P^{\dagger}\xi P$. The trilinear terms  proportional
to $\tilde{\lambda}_6$ and $\tilde{\lambda}_7$, are often dropped by requiring
the discrete symmetry $\Delta \to -\Delta$ and $\chi \to -\chi$. This symmetry
then has to be implemented also on the Yukawa term
\begin{equation} \label{e11}
{\cal L}_{ll\Delta}=ih_{ab}{\psi_{aL}^T}C\tau_2{\tau_i \over 2}\psi_{bL}
\Delta^i_P+ {\rm h.c.},\, \, \, \, \, \, \, \psi_i=\left(\begin{array}{c}
\nu_i \\
\l_i
\end{array}
\right)
\end{equation}
resulting in a neutrino spectrum with  one Weyl and one massive Dirac
neutrino. Hence, in principle, the model with the discrete symmetry has
different physical consequences as compared to the more general model
of eq. (\ref{e10}) as far as the neutrino mass spectrum is concerned.

Diagonalising  the
mass matrix of the scalar sector, in general one finds, after duly absorbing
the Goldstone bosons as longitudinal components of gauge fields, a 5-plet,
${H_5}^{++,+,0,-,--}$, a 3-plet, ${H_3}^{+,0,-}$ and two singlets, ${H_1}^0$
and ${H_1}'^0$, as the physical states. The different multiplets
charaterise their respective transformation properties under the custodial
$SU(2)$. The members of the $H_5$-plet are given by (for further details on the
particle spectrum and couplings see \cite{gunion})
\begin {eqnarray} \label{e12}
 H_{5}^{++}& =& \Delta^{++} \nonumber \\
 H_{5}^{+}& =& (\Delta^{+} -  \chi^{+})/\sqrt{2}\nonumber \\
 H_{5}^{0}& =& (2 \chi^{0} - \sqrt{2} \Delta^{0})/\sqrt{6}
 \end {eqnarray}
Clearly, none of them have any overlap with the components of the
doublets $\bf{\phi}$, and as such they do not couple to fermions at the
tree-level in contrast to the case of the physical charged Higgs $H^{'+}$ of
eq. (\ref{e5}) of the earlier model. The mass of the 5-plet members is given by
\begin{equation} \label{e13}
M_5^2=8\tilde{\lambda}_4w^2-3\tilde{\lambda}_5v^2-{1 \over 2}\tilde{\lambda}_6
{v^2 \over w}
\end{equation}

  One constraint on the above scenario arises from the lepton-lepton
couplings that the complex Y=2 triplet possesses (see eq. (\ref{e11})). The
mass thus acquired by the electron neutrino will, for example with
diagonal $h_{ab}$, be given by

\begin {equation} \label{e14}
M_{\nu_e} = h_{ee} {s_{H} \over 2} {M_W \over g}
\end {equation}
with
\begin{equation} \label{ex1}
s_{H}={2w \over \sqrt{v^2 + 4w^2}}
\end{equation}
and $g$ is the $SU(2)$ coupling constant.
The experimental constraints from neutrinoless double beta-decay
imply that $M_{\nu_e} < 1 eV$. This means that either the doublet-triplet
mixing angle or the $\Delta L=2$ Yukawa coupling is restricted to very
small values. Since in this model $\rho=1$ at tree level
a large mixing angle $s_H$ is {\it apriori} not excluded. A large vev $w$ will
imply of course larger contribution to the gauge boson mass coming
from the triplet sector.
We will comment on this further at the end of the paper. It has also
been shown that so far as scalar interactions are concerned,
one can  maintain the equality of vev's at higher orders.
However, no way has yet been found to prevent the custodial
symmetry from being broken at the loop-level
by $U(1)$ gauge couplings. This means that fine-tuning is required to make
such a model work. In any case, as has been discussed in ref. \cite{gunion},
the degree of such fine-tuning required is not higher than that involved
in connection with the naturalness problem within the standard model itself
(see also \cite{kunszt} in this connection).

As has been mentioned before, in both the models
discussed above the $H^+W^-Z$ coupling exists
at the tree level. The lagrangians are
\begin{eqnarray} \label{e15}
{\cal L}^{(1)}_{HWZ}&=&-{g M_W \over \cos \theta_W} s_{H'}c_{H'}H^{'+}W^-_{\mu}
Z^{\mu} +{\rm h.c.}\nonumber \\
{\cal L}^{(2)}_{HWZ}&=&-{g M_W \over \cos \theta_W} s_{H}H^{+}_5W^-_{\mu}
Z^{\mu} +{\rm h.c.}
\end{eqnarray}

  We now focus on the production of ${H}^{\pm} (H^{'\pm} $  or
$H_5^{\pm})$ via the above interactions. First, there is the
s-channel process $e^{+}e^{-} \longrightarrow Z^* \longrightarrow W H^{\pm} $
which will dominate at lower energies. Here we concentrate upon the final
states consisting of the leptonic decay products of the W, i.e.  on
$e^{+}e^{-} \longrightarrow H^{\pm} l \nu_{l}$. In such cases, the $e \nu_e $
final state also receives  contributions from a t-channel diagram with
the $H^{\pm} $ emitted from the propagator. As we shall see later,
for high values of the centre-of-mass energy, this latter diagram
is dominant.

  Fig. 1 shows the cross-sections for $H^{\pm}l{\nu_l}$-production plotted
against $M_{H^{\pm}}$ for different values of $\sqrt{s}$. The contributions due
to
the electronic and muonic final states are added. The curves correspond
to $s_H =1$ and $c_{H'} s_{H'} = 1$ for the two cases of $H_5^{\pm}$,
$H'^{\pm}$ respectively.  The cross-sections for various values
of the mixing angles can be read off by multiplying by the appropriate value of
$s_H^2$($s^2_{H'} c^2_{H'} \simeq s^2_{H'}$) . It is
straightforward to see from the graphs that in the LEP-II case (thin solid
line), assuming an
integrated luminosity of $10^{39} cm^{-2}$ per year there will be
a few hundreds of events for $s_H =1$
upto at least $M_{H_5} = 110 \, GeV$. If now one
has only the $Y=2$ triplet, the restriction from the ${\rho}$-parameter
allowes a maximum  $s_{H}^2$ of 0.009. This leaves us with about  2 events
per year. In such a case, the only reasonable chance of observing this
process exists in a higher-energy $e^{+} e^{-}$ machine. As can be seen from
the plot (dotted line), such a machine with a luminosity
of $50 fb^{-1}/ {\rm year}$ \cite{nlc} can lead to few tens of events upto
 $M_{H'} \simeq 300$ GeV even with a value of $s_{H'}$ well
within the limits imposed by $\rho$. For sake of completeness
we also show the expected cross section for higher values of $\sqrt{s}$
 ($1$ TeV (dashed line) and $2$ TeV (thick solid line)). On the other hand,
since this restriction gets relaxed if one presupposes a complex
and a real triplet, at  LEP-II itself one can investigate
a large area of the ${\rho}-s_H$  parameter space in the latter scenario.

It is instructive to note here that while the s-channel diagram dominates at
lower beam-energies ($\sqrt{s} = 200 $ GeV) and hence the cross--sections for
the electronic and muonic final states are not too different (shown in fig. 2
by the solid and dashed line, respectively), at higher
energies the t-channel clearly dominates. For example, the cross--section
for the electronic final state (shown by solid line) is at least a factor 5
larger than that for muonic final state (dashed line) already at
$\sqrt{s} = 500$ GeV for $ M_{H^\pm} \leq 150$ GeV.
For higher values of $\sqrt{s} $ the cross--sections is
almost completely dominated by the process $e^+ e^- \to e \nu_e H^{\pm}$

     The signals of the triplet scalars thus produced also of course
depend in a
rather crucial manner on their subsequent decay channels. For the
(complex+real) case, as has already been mentioned, the charged scalars
$H_5^{\pm}$ do not have tree-level interactions with fermions. Possibilities
of observing them through loop-induced decays into fermion pairs
have already been studied \cite{gunion}. However, we
would like to point out here that there also exists the tree-level decays
into four-fermions mediated by a W and a Z coupling with $H_5$. So
long as $M_{H_5} < M_W$, both the W and the Z in this decay are virtual.
We have explicitly calculated such decay widths using
methods described in ref. \cite{weiler} modified appropriately. The
width up to mixing angles is given by
\begin{eqnarray}
\label{e16}
\Gamma(H^+ \to W^*Z^* &\to & f_1 \bar{f}_1 f_2 \bar{f'}_2)={g^6 m_W^2 \over
2^9 \times 9 \times M_{H^+}^3 \times (2\pi)^5 \times \cos^4 \theta_W}
(g_V^2 + g_A^2) \nonumber \\
&\times &\int_0^{M^2_{H^+}}dQ_1^2 \int_0^{(M_{H^+}-\sqrt{Q_1^2})^2}dQ_2^2
\lambda^{1/2}(M_{H^+}^2, Q_1^2, Q_2^2) \nonumber \\
&\times &[8Q_1^2 Q_2^2 + (M_{H^+}^2-Q_1^2 -Q_2^2)^2]B_Z(Q_1^2)B_W(Q_2^2)
\end{eqnarray}
where $g_V=T_{3L}^{f_1}-2Q^{f_1}\sin^2 \theta_W$, $g_A=-T_{3L}^{f_1}$,
$B_V(Q^2)=[(Q^2-M_V^2)^2 + M_V^2 \Gamma_V^2]^{-1}$ and $\lambda(x,y,z)$
is the kinematical triangle function.
We show the decay widths into this channel for leptonic final states
(with $s_H =1$) as a function of $M_{H_5}$ by the dashed line in fig. 3.
Here we have summed over the muons and the electrons  in the final state.
For purposes of comparison we also show by the dotted line
one of the representative results for the  loop
induced  partial width $\Gamma (H_5^+ \to c \bar s)$ taken from
ref. \cite{gunion} (also drawn there  for $s_H=1$). The figure clearly
shows that indeed the partial decay width for our four-fermion  modes
are of similar orders for $M_{H_5} \sim 50 GeV$, and completely dominate
for higher values of $M_{H_5}$.
\footnote{It should be pointed out here that even for $M_{H_5}$ as small as 25
GeV, the four-fermion width (summed over leptons and light quarks) is somewhat
larger than the loop induced $\Gamma(H_5 \to c \bar s)$. If we further
remember that the loop induced $\Gamma(H_5 \to \tau \nu_{\tau})$ is even
smaller
than that for $c \bar s$ final state \cite{gunion}, this implies that the
normal search strategies for a charged Higgs at LEP might have missed such
a charged Higgs.} When $M_{H_5}$ becomes greater than $M_W$,
the loop-induced $W\gamma$ channel also opens up for the $H_5$. However, if we
take the numbers for the partial decay width for this channel given in
\cite{ray} as a very rough guideline, we find that the decay mode into the four
fermions (via two gauge bosons (real or virtual)) does indeed dominate.
Consequently, the branching ratio for  decays into a pair of weak gauge-bosons
(WZ) remains close to 100 per cent over most of the parameter
space under scrutiny.  The 4-lepton ($ l \nu_l l^{'+} l^{'-}$) channel then has
a healthy branching ratio of $\sim 1.4 \%$ The corresponding signal
is practically free from standard model backgrounds, and thus it should be
considered as the principal technique in looking for triplet Higgs bosons
(in this model) produced in $e^{+} e^{-}$ collider experiments. Thus using this
clean lepton channel, for $s_H = 1$, LEP-200 (NLC) can have a discovery range
upto $M_{H_5} = 100 (350) $ GeV. The signal with four jets in the final
state might seem hopeless at first
glance when one thinks of multijets coming from gauge boson pair production
and their decays. However, it should be remembered that the invariant mass of
the four jets will be quite different from $\sqrt{s}$ and hence even the
four jet final states can be used increasing the useful branching ratio to
$\sim 50\%$. This would increase the discovery range of $H_5^{\pm}$ even
further since the width into four quarks final state will be a factor $\sim 35$
higher than the four lepton channel. Hence, the discovery range at LEP-200
(NLC) gets extended to 125 (425) GeV, for $s_H = 1$. It should also be pointed
out that the $e^-/e^+$ coming from the reaction $e^+ e^- \to  e^{+}
H_5^{-} \bar \nu_e  (c.c.) $  is likely to be lost in the beam pipe.

For the case of model 1 (i.e. $H^{'+}$) of course the two-fermion decay-modes
can occur at tree level just like the $WZ$ decay and are thus not loop
suppressed though suppressed by the mixing angle $s_{H'} c_{H'}$. As can be
seen from $\Gamma (H^{'+} \to c \bar s)$ given by the solid line in fig. 3
(and dotted line in  fig. 4) this decay mode will dominate  over  the
four-lepton decay mode upto $M_{H^{'+}} = 150$ GeV.
However the width for the 4-quark decay mode
(mediated by $WZ$ (real or virtual)), shown by dashed line in
fig. 4,   will be comparable  to the two quark decay mode for
$M_{H^{'+}} \geq 120 $ GeV as can be seen from fig. 4.  For still heavier
$H^{'+} $ (once the $WZ$ channel is allowed kinematically and the $H^{'+} \to t
\bar b$ opens up) these two decay modes (shown by the thin and thick solid line
respectively in fig. 4) will take over and eventually the
branching ratio into the clean 4-lepton channel will be $\sim 0.7 \%$ level.
Since all the relevant decays occur only through the mixing the actual value of
the mixing angle $s_{H'}$ is immaterial for gauging the relative strengths of
different channels. Of course, the statements about the tree-level decays into
$WZ, t \bar b$ and the four jet channels, made above are true for the case
of $H_{5}^{+}$ as well.

   At the end, we would like to make a comment regarding the production of
doubly charged Higgs bosons which form an integral part of the triplet models
discussed here. If one considers the process
$e^{+} e^{-} \longrightarrow H^{++} l^{-} l^-$ (or its charge
conjugate), the production rate is proportional to the square of
the $H^{++}$-lepton-lepton coupling strength. This strength is bounded above
from the non-observation of neutrinoless double beta decay.
As is evident from eqs. (\ref{e14},\ref{ex1}), the maximum value of the
coupling is thus inversely proportional to
the quantity $s_H$. A rather model-independent limit can be obtained from
indirect contributions of $\Delta^{++}$ to Bhabha scattering (t-channel
exchange) \cite{swartz}
\begin{equation} \label{e17}
{h^2_{ee} \over M^2_{++}} < 1.9 \times 10^{-5}\, \, {\rm GeV}^{-2}
\end{equation}
Therefore, for very small $s_H$, the production
cross-section for a single doubly-charged Higgs becomes less restricted.
In the extreme cases, $h_{ee}$ very small or $w/v$ very small, there is an
interesting dichotomy which can be used for experimental studies: either
the $HWZ$ coupling is sizeable or the coupling strength of $\Delta^{++}$ to
leptons is non-negligible.
A detailed exploration of the signal of a such a singly-produced doubly
charged Higgs is thus advisable. Although this production mechanism has
been studied previously \cite{petrarca}, some relevant diagrams have been left
out, which are potentially important for a high-energy $e^{+} e^{-}$ machine.
A study of the full process is currently under way.

  In conclusion, The tree level $HWZ$-coupling in the triplet model is found
to be  a rather interesting way of either uncovering or ruling out such
a scenario.
 For a model with only complex triplets, LEP-II can have at best a marginal
 glimpse of the allowed region of the parameter space, and machines with
 higher energy and luminosity are required for a closer survey. On the other
 hand, with one complex and one real triplet a considerably
 large region of the allowed parameter space is likely to
 come within the purview of LEP-II, with conspicuous and testable signals.
\vskip 2truecm

{\bf Acknowledgements}. We thank J.W.F.Valle, A.S.Joshipura and S.Rindani for
useful discussions.  We also thank the Department of Science and Technology
(India) and
the organisers of WHEPP-III held at Madras (India) where this work was started.
R.G. and B.M. would like to thank the International Centre
of Theoretical Physics for hospitality where this work was completed. The work
of R.G was supported in part by the Council for Scientific and Industrial
Research under a research grant. M.N. wishes to
acknowledge the financial support by the HCM program under EEC contract number
CHRX-CT 920026.

\newpage
\noindent{\bf Figure Captions:}

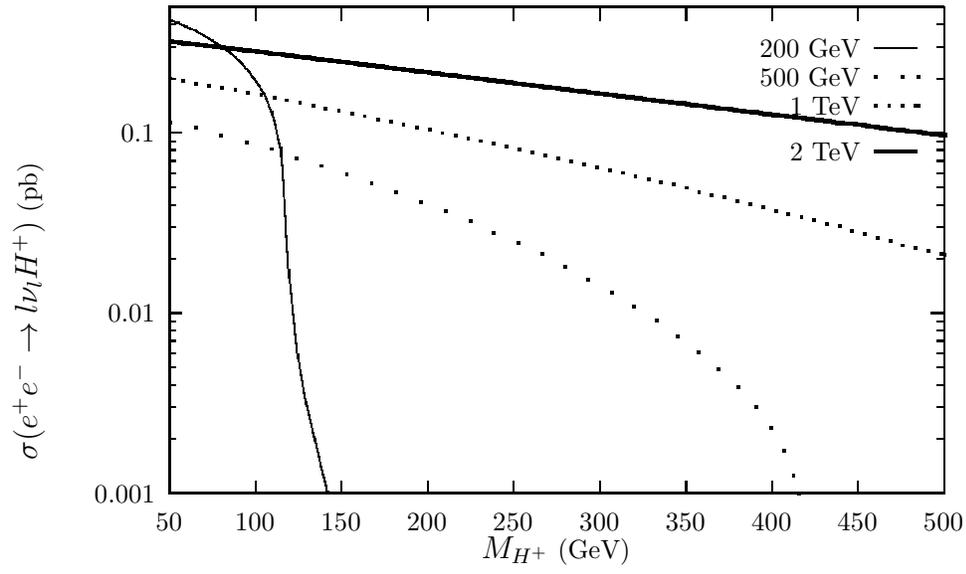
\noindent {\bf Fig. 1.} $\sigma(e^+ e^- \to l \nu_l H^{\pm}) $ (pb) as a
function of
$M_{H^+}$ (for $s_H (s_{H'}c_{H'}) = 1)$ for different values of
$e^+e^-$ center of mass energies: the thin
solid, dotted, dashed and thick solid line correspond to $\sqrt{s}= $ 200 GeV,
500 GeV, 1 TeV and 2 TeV respectively. Cross-sections for both charges of the
Higgs and muonic as well as electronic channels in the final state are added.

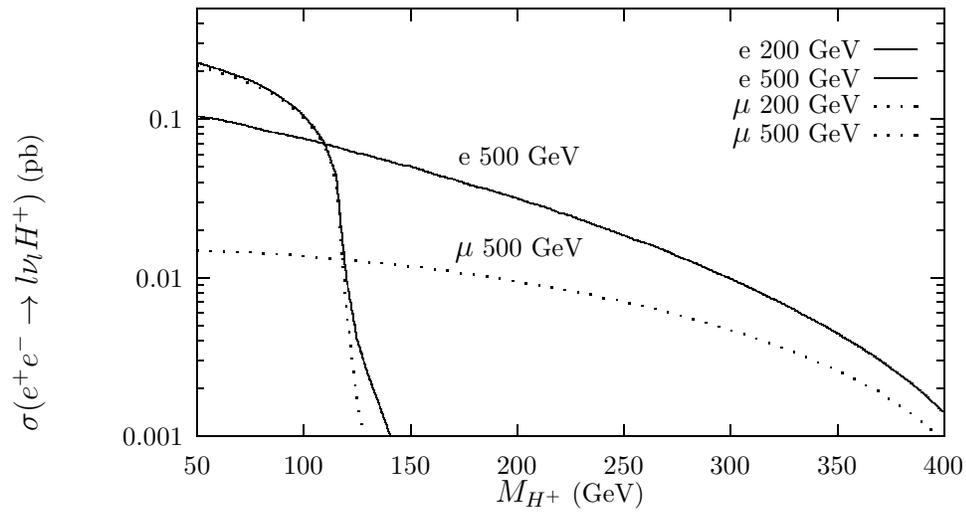
\noindent {\bf Fig. 2.} $\sigma(e^+ e^- \to e \nu_{e} H^{\pm}) $ (solid line)
and  $\sigma(e^+ e^- \to \mu \nu_{\mu} H^{\pm}) $ (dashed line) for
$\sqrt{s} =$ 200 GeV and 500 GeV, for same values of the mixing angle as in
fig. 1 and again summed over the charge  of the Higgs.

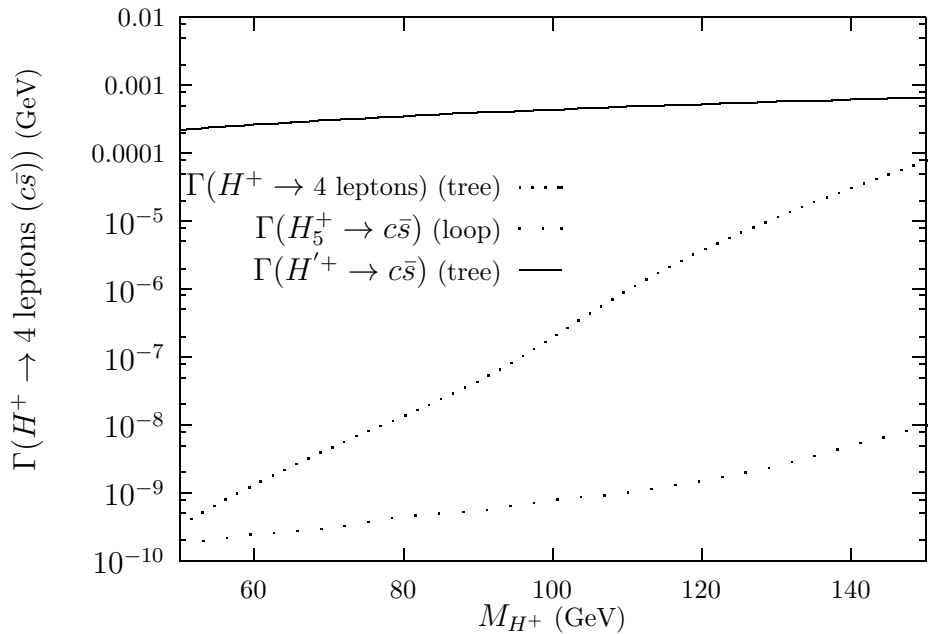
\noindent {\bf Fig. 3.} The different possible two body  decay widths
($H^{+}_5 \to c \bar s$)at loop level (dotted line), tree level (solid line)
 and the tree level leptonic four body decay width
 $\Gamma (H^{+}_{5} \to W^* Z ^* \to 4 \ {\rm leptons})$  (dashed line) for
the charged Higgs. Again the mixing angles are put equal
to 1 as in fig. 1. The $\Gamma (H^+_5 \to c \bar s )$ (loop) has been taken
from
ref. \cite{gunion}. The four body decay width is summed over electrons and
$\mu 's $ in the final state. For four quark final state the numbers are
obtained by multiplying the dashed figure by $\sim 35$.

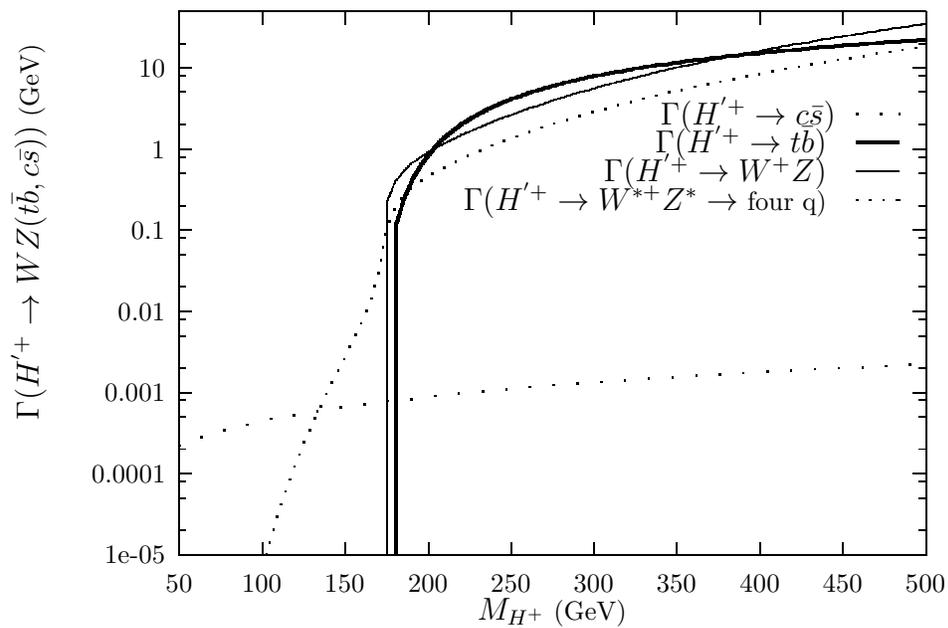
\noindent {\bf Fig. 4.} The different possible tree level decay widths
($H^{'+} \to c \bar s$) (dotted line),
 $\Gamma (H^{'+} \to W^* Z ^* \to 4 \ {\rm quarks})$  (dashed line),
$\Gamma ( H^{'+} \to W^+ Z )$ (thin solid line) and
$\Gamma ( H^{'+} \to t \bar b )$ (thick solid line) for
the charged Higgs. Again the mixing angles are put equal
to 1 as in fig. 1. $m_b$ has been neglected.

\newpage
\pagestyle{empty}
% GNUPLOT: LaTeX picture
\setlength{\unitlength}{0.240900pt}
\ifx\plotpoint\undefined\newsavebox{\plotpoint}\fi
\sbox{\plotpoint}{\rule[-0.200pt]{0.400pt}{0.400pt}}%

\begin{center}
{\large Fig. 4}
\end{center}

\newpage
\begin{thebibliography}{99}
\bibitem{2hd}
The standard reference for extended Higgs sectors is J.~Gunion, H.~E.~Haber,
G.~Kane and S.~Dawson, `The Higgs Hunter's Guide', Addison-Wesley Publishing
Company 1990; for a recent study of a two Higgs doublet model see for example
A.~Pilaftsis
and M.~Nowakowski, Int.~J.~Mod.~Phys.~{\bf A9} (1994) 1097

\bibitem{tri}
H.~M.~Georgi, S.~L.~Glashow and S.~Nussinov, Nucl.~Phys.~{\bf B193} (1981) 297

\bibitem{solar}
see e.g. S.~M.~Bilenky and S.~T.~Petcov, Rev.~Mod.Phys.~{\bf 59} (1987) 671;
J.~W.~F.~Valle, Prog.~Nucl.~Phys.~{\bf 26} (1991) 91

\bibitem{rho}
Review of Particle Properties, Phys. Rev. {\bf D50} (1994).
\bibitem{georgi}
H.~Georgi and M.~Machacek, Nucl.~Phys.~{\bf B262} (1985) 463

\bibitem{chano}
S.~Chanowitz and M.~Golden, Phys.~Lett.~{\bf B165} (1985) 105

\bibitem{dicus}
R.~Vega and D.~Dicus, Nucl.~Phys.~{\bf B329} (1990) 533

\bibitem{gunion}
J.~F.~Gunion, R.~Vega and J.~Wudka, Phys.~Rev.~{\bf D42} (1990) 1673; {\bf D43}
(1991) 2322

\bibitem{hwz}
J.~A.~Grifols and A.~Mendez, Phys.~Rev.~{\bf D22} (1980) 1725; M.~C.~Peyranere,
H.~E.~Haber and P.~Irulegui, Phys.~Rev.~{\bf D44} (1991) 191

\bibitem{bisw}
B.~Mukhopadhyaya, Phys.~Lett.~{\bf B252} (1992) 123
\bibitem{ronc}
G.~B.~Gelmini and M.~Roncadelli, Phys.~Lett.~{\bf B99} (1981) 411

\bibitem{stri}
J.~R.~Espinosa and M.~Quiros, Nucl.~Phys.~{\bf B384} (1992) 113
\bibitem{kunszt}
P.~Bamert and Z.~Kunszt, Phys.~Lett.~{\bf B306} (1993) 335

\bibitem{nlc}
see, e.g., M.~Drees and R.~M.~Godbole, Z. Phys. {\bf C 59} (1993) 591;
P. Chen, T. L. Barklow and M.E. Peskin, Phys. Rev. {\bf D49} (1994) 3209.
\bibitem{weiler}
H.~Pois, T.~J.~Weiler and T.~C.~Yuan, Phys.~Rev.~{\bf D47} (1993) 3886
\bibitem{ray}
S.~Raychaudhuri and A.~Raychaudhuri, Phys.~Rev.~{\bf D50} (1994) 412.
\bibitem{swartz}
M.~L.~Swartz, Phys.~Rev.~{\bf D40} (1989) 1521
\bibitem{petrarca}
M.~Lusignoli and S.~Petrarca, Phys.~Lett.~{\bf B226} (1989) 397
\end{thebibliography}
\end{document}